\documentclass[letterpaper]{article}
\usepackage{uai2019}
\usepackage[margin=1in]{geometry}

\usepackage{times}
\usepackage[utf8]{inputenc}
\usepackage{mathtools}
\usepackage{amsfonts}
\usepackage{booktabs}
\usepackage{algorithm}
\usepackage{algpseudocode}
\usepackage[colorlinks=true,citecolor=blue,urlcolor=blue]{hyperref}

\newcommand{\abs}[1]{\left|#1\right|}
\newcommand{\et}{\;\mathrm{and}\;}
\newcommand{\Real}{\mathbb{R}}
\newcommand{\STAB}{\mathrm{STAB}}
\renewcommand{\TH}{\mathrm{TH}}
\newcommand{\chull}{\mathrm{Conv}}
\newcommand{\PA}{\mathrm{PA}}
\newcommand{\pa}{\mathrm{pa}}

\title{Exclusivity graph approach to Instrumental inequalities}
\author{
{\bf Davide Poderini\textsuperscript{1}, 
     Rafael Chaves\textsuperscript{2,3}, 
     Iris Agresti\textsuperscript{1}, 
     Gonzalo Carvacho\textsuperscript{1}, 
     Fabio Sciarrino\textsuperscript{1}}\\
\textsuperscript{1}{\it
Dipartimento di Fisica,
Sapienza Universit\'a di Roma, 
Piazzale Aldo Moro 5, I-00185 Roma, Italy}\\
\textsuperscript{2}{\it
International Institute of Physics, 
Federal University of Rio Grande do Norte, 
59070-405 Natal, Brazil}\\
\textsuperscript{3}{\it
School of Science and Technology, 
Federal University of Rio Grande do Norte, 
59078-970 Natal, Brazil}\\
}

\begin{document}

\maketitle

\begin{abstract}
Instrumental variables allow the estimation of cause and effect relations even
in presence of unobserved latent factors, thus providing a powerful tool for any
science wherein causal inference plays an important role. More recently, the
instrumental scenario has also attracted increasing attention in quantum
physics, since it is related to the seminal Bell's theorem and in fact allows
the detection of even stronger quantum effects, thus enhancing our current
capabilities to process information and becoming a valuable tool in quantum
cryptography. In this work, we further explore this bridge between causality and
quantum theory and apply a technique, originally developed in the field of
quantum foundations, to express the constraints implied by causal relations in
the language of graph theory. This new approach can be applied to any causal
model containing a latent variable. Here, by focusing on the instrumental
scenario, it allows us to easily reproduce known results as well as obtain new
ones and gain new insights on the connections and differences between the
instrumental and the Bell scenarios. 
\end{abstract}

\section{INTRODUCTION}
Inferring  whether a variable $A$ is the cause of another variable
$B$ is at the core of causal inference \cite{Mooij}. However, unless
interventions are available \cite{pearlbook}, one can cannot exclude that
observed correlations between $A$ and $B$ are due to a latent common
factor, thus hindering any causal conclusions. To cope with that,
instrumental variables (IV) have been introduced \cite{pearl1995,
bonet2001}. Under the assumption that they are independent of any
latent common factors $\Lambda$, IV can be used to put non-trivial
bounds on the causal effect between $A$ and $B$. However, first,
one has to guarantee that an appropriate instrument (fulfilling a
set of causal constraints) has been employed, which is precisely
the goal of the so-called instrumental tests \cite{pearl1995,
bonet2001,Ramsahai2012,Kedagni2017}. Their violation, at least in
classical physics, is an unambiguous proof that some of the causal
assumptions underlying the instrumental causal structure are not
fulfilled, that is, one should identify and use another instrumental
variable.

Instrumental tests have firstly been introduced in econometrics
\cite{Wooldridge2015} and further explored by Pearl \cite{pearl1995},
in the form of inequalities providing a necessary condition for a
given observed probability distribution to be compatible with the
instrumental causal structure. Following that, Bonet \cite{bonet2001}
introduced a general framework also followed in \cite{Ramsahai2012},
showing that the instrumental correlations define a polytope, a convex
set from which the non-trivial boundaries are precisely the instrumental
inequalities. Bonet's framework allows for the derivation of new
inequalities as well as proving general results, for instance, the fact
that if variable $A$ is continuous, no instrumental test exists. However,
two main drawbacks arise. First, the systematic derivation of new
inequalities quickly becomes unfeasible as the variables' cardinalities
increase. Second, as recently shown, in quantum physics, violations of
the instrumental tests are possible even though the whole process is
indeed subjected to an instrumental causal structure \cite{chaves2018,
himbeeck2018}. In the quantum case, instrumentality violations witness
the presence of quantum entanglement as the latent factor and prove
a stronger form of quantum non-locality compared to the famous Bell's
theorem \cite{chaves2018}. As a consequence, typical bounds on the causal
influence of $A$ into $B$ have to be reevaluated and reinterpreted in
the presence of quantum effects.

Our aim in this paper is to provide a novel and complementary framework
to the analysis of instrumental tests, which also addresses the two
drawbacks mentioned above. The proposed method is based on a graph
theoretical approach introduced in the foundations of quantum physics
to analyze the possible correlations obtained in quantum experiments
\cite{cabello2014,rabelo2014}. This method allow us to reproduce the
classical results by Bonet and to straightforwardly generalize them in
the quantum scenario. It also offers an easy and general way -- valid
for any causal scenario involving a single latent variable -- to check
for the incompatibility between the quantum and classical descriptions.

The paper is organized as follows: first we provide the necessary
background for our work, describing the instrumental and Bell scenarios
from both classical and quantum perspectives and introducing the
exclusivity graph approach. We then show the versatily of our framework
by rederiving and generalizing known results in the literature, obtaining
new instrumental inequalities that hardly could be found by standard
means and offering new insights about the similarities and differences
between instrumental and Bell inequalities.

\begin{figure}[t]
    \centering
    \includegraphics[width=.8\columnwidth]{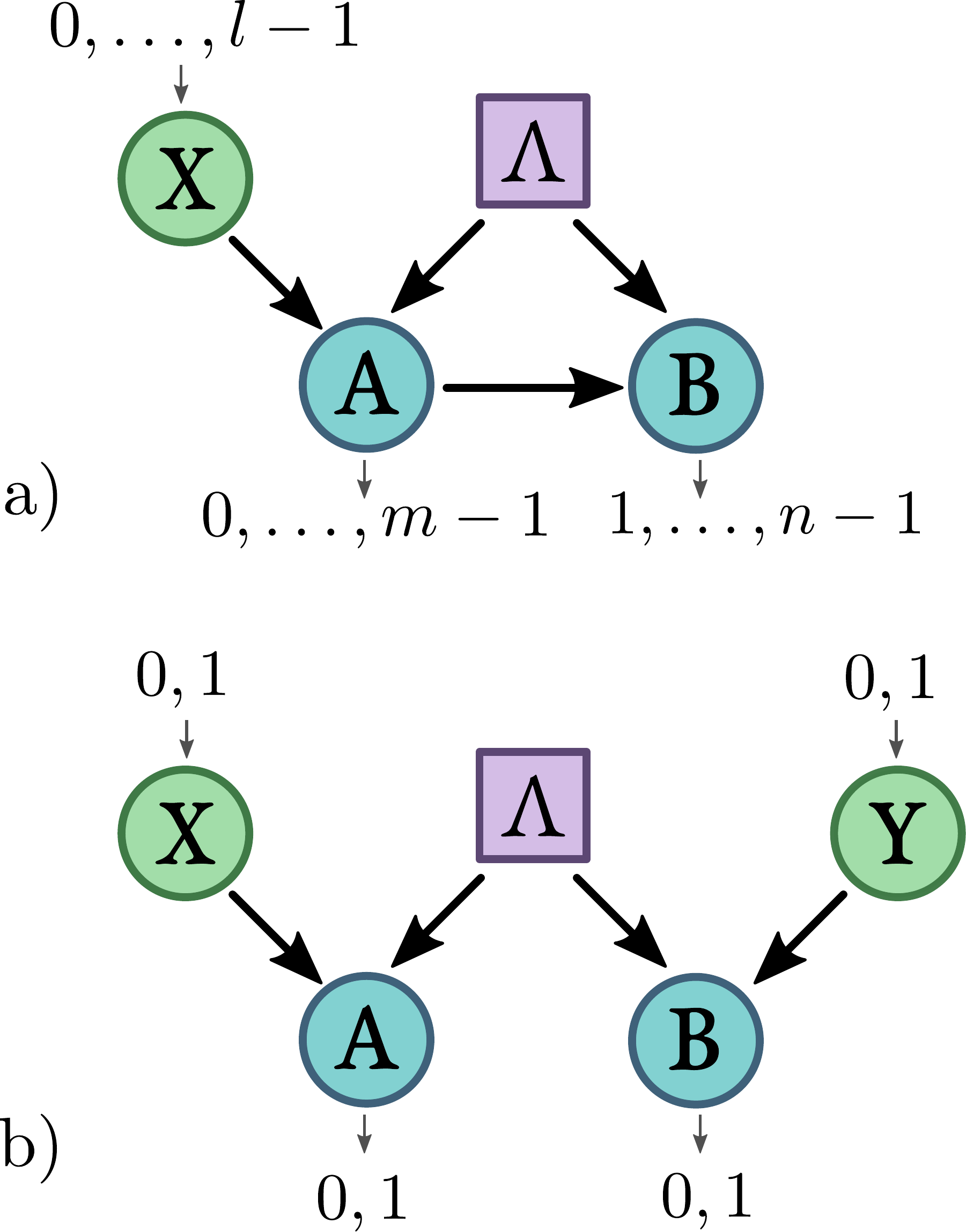}
        \caption{
    Directed acyclic graph (DAG) representation for
    \textbf{a)} a general
    Instrumental scenario, with $l$ possible values for the random variable $X$
    and $m,n$ possible outcomes for $A$ and $B$ respectively, and for
    \textbf{b)} the CHSH scenario \cite{CHSH} where all the
    variables $X,Y,A$ and $B$ can only take two possible values.
}
    \label{fig:chshinstdag}

\end{figure} 

\section{INSTRUMENTAL VARIABLES, ESTIMATION OF CAUSAL INFLUENCES AND A NEW FORM OF QUANTUM NON-LOCALITY}

We represent causal relations via directed acyclic graphs (DAG), where the
nodes represent random variables interconnected by directed edges (arrows)
accounting for their cause and effect relations \cite{pearlbook}.  A set
of variables  $\left( X_1,\dots, X_n \right)$ form a Bayesian network
with respect to the graph if every variable $X_i$ can be expressed as
a function of its parents $\PA_i$ and potentially an unobserved noise
term $U_i$, such that $U_i$ are jointly independent. This implies that
the probability distribution of such variables should have a Markov
decomposition
\footnote{Uppercase letters label variables and lowercase label the values
taken by them, for instance, $p(X_i =x_i, X_j = x_j) \equiv p(x_i, x_j)$.}
\begin{equation}
p(x_1,\dots,x_n)= \prod_{i=1}^{n} p(x_i \vert \pa_i).    
\end{equation}
Importantly, a DAG typically implies non-trivial constraints over
the probability distributions that are compatible with it. That is,
simply from observational data and without the need of interventions,
one can test whether some observed correlations are incompatible with
some causal hypotheses.

Within this context, an important DAG is that corresponding to the instrumental
scenario (see Fig.\ref{fig:chshinstdag}-a). Following the Markov decomposition, any
empirical data encoded in the probability distribution $p(a,b \vert x)$ and
compatible with the instrumental causal structure can be decomposed as
\begin{equation}
p(a,b \vert x) = \sum_{\lambda} p(a\vert x,\lambda) p(b\vert a,\lambda)p(\lambda).
\end{equation}
Two causal assumptions are employed to arrive to the above
decomposition. i) The first assumption $p(x,\lambda)=p(x)p(\lambda)$,
implies the independence of the instrument and the common ancestor. ii)
The second assumption requests that, even though $X$ and $B$ can be
correlated, all these correlations are mediated by $A$. In other terms,
there is no direct causal influence between $X$ and $B$: $p(b\vert
x,a,\lambda)=p(b\vert a,\lambda)$.

The instrumental variables have been originally introduced to estimate
parameters in econometric models of supply and demand \cite{economic}
and, since then, have found a wide range of applications in various other
fields \cite{economic2, economic3}. To illustrate its power, consider
that variables $A$ and $B$ are related by a simple linear equation,
i.e. $B=\gamma A +\Lambda$, where $\Lambda$ may represent a latent
common factor. By assumption, the instrumental variable $X$ should be
independent of $\Lambda$, thus implying that the causal strength can
be estimated as $\gamma= \mathrm{Cov}(X,B)/\mathrm{Cov}(X,A)$ where
$\mathrm{Cov}(X,A)$ is the covariance between $X$ and $A$. Strikingly,
one can estimate the causal strength even without any information about
the latent factor $\Lambda$. More generally, and without assumptions
about the functional dependence among the variables, the empirical data
encoded in the probability distribution $p(a,b \vert x)$ can also be
used to bound different quantifiers of causality between $A$ and $B$
\cite{pearlbook,Janzing2013}.

Clearly, however, to draw any causal conclusions, first it is necessary to
certify that one has a valid instrument. This is achieved via instrumental
inequalities, first introduced by Pearl \cite{pearl1995}. If we allow the
variables $X$, $A$, $B$ to take the values in the range  $x=1,\dots,l$,
$a=1,\dots,m$ and $b=1,\dots,n$ Pearl showed that the instrumental causal
structure implies (independently of any assumption about the functional
dependence among the variables) that
\begin{equation} 
    \sum_{j=0}^{n} P(ij|k(i,j)) \le 1,
    \label{eq:pearl_ineq}
\end{equation}
for all $i \in {1,\ldots, m}$ and for all the possible functions $k(i,j)$
where $p(i,j|k) = p(a=i,b=j\vert x=k)$.

Extending these results, Bonet \cite{bonet2001} provided a general
geometric framework for the derivation of instrumental inequalities.
Instrumental correlations define a convex set, a polytope described
by finitely many extremal points, or alternatively by a finite number
of facets, among which, the non-trivial are precisely the instrumental
inequalities. In particular, considering the case $(l,m,n) = (3,2,2)$,
it was proven that there are two inequivalent classes of instrumental
inequalities (those not obtained from each other by permuting the
labels of $i,j$ and $k$). One class corresponding to Pearl's inequality
\eqref{eq:pearl_ineq} and the other given by
\begin{multline}
    P(0 0 | 0) + P(1 1 | 0) + 
    P(0 0 | 1) +\\+ P(1 0 | 1) + 
    P(0 1 | 2) \le 2.
    \label{eq:bonet_ineq}
\end{multline}

All these conclusions and results, however, rely on a classical
description of causal and effect relations (implicitly) invoking the
realism assumption: the probabilities of a given measurement have well
defined values even if such measurements are not performed. However,
since Bell's theorem \cite{bell1964} we know that this do not apply to the
world governed by quantum mechanics, thus implying that standard causal
models, even if augmented with latent variables, are not enough to explain
quantum phenomena. Bell's theorem relies on the causal structure shown in
Fig.~\ref{fig:chshinstdag}-b, similar to the instrumental one but with two
crucial causal differences: i) variable $A$ has no causal influence over
$B$ and ii) $B$ has its own instrument $Y$ and thus the correlations are
encoded in a probability distribution $p(ab \vert xy)$. This has motivated
the question of whether many of the cornerstones in causal inference have
to be re-evaluated or reinterpreted in the presence of quantum effects
\cite{ried2015,Costa2016,Wolfe2016,gonz1,gonz2}. Indeed, as recently
shown \cite{chaves2018}, violations of the instrumental tests are
possible even though the causal constraints underlying the instrumental
scenario are fulfilled. As shown in the experimental implementation
of the instrumental test \cite{chaves2018}, this is possible due to
the presence of quantum entanglement that precludes the explanation of
the data via a latent common factor. Interestingly, every probability
distribution violating the simplest possible Bell inequality, known as
Clauser-Horne-Shimony-Holt (CHSH) inequality \cite{CHSH}, can after some
post-processing also violate Bonet's inequality \cite{himbeeck2018}. As we
will see, the graph-theoretical approach will allow us a more systematic
understanding of the similarities and differences between the Bell and
instrumental scenarios.

Altogether, this shows the necessity of a new unifying framework, not
only considering what are the classical instrumental correlations but
as well the ones achievable if the underlying latent factor might have
a quantum origin. In the following we will achieve that by proposing a
graph-theoretical approach to analyze the instrumental inequalities.

\section{THE EXCLUSIVITY GRAPH APPROACH}
The graph-theoretical approach we propose here, was initially developed
for the study of non-contextual inequalities \cite{cabello2014} as well
as Bell non-locality scenarios \cite{acin2015}.  The purpose of this
method is to easily obtain constraints on the probability distribution,
in the same spirit of the Pearl's and Bonet's inequalities, for classical
and quantum systems. In the following we will have a set of random
variables $A_1,\ldots,A_N$ representing the outcomes of measurements
performed on our physical system, and another set $X_1,\ldots,X_M$, a
number of measurement settings that can be chosen by the experimenter,
which serve the same purpose of the instrument in the IV scenario. In the
exclusivity graph formalism, every possible event, i.e. every possible
set of measurement outcomes $a_1,\ldots, a_N$ corresponding to given
measurement settings $x_1,\ldots,x_M$ , is associated to a vertex
in a (undirected) graph $G = (V, E)$.  Two vertices $u, v \in V$ are
connected by an edge $(u,v) \in E$ if and only if they are exclusive,
i.e., if there is a measurement/instrument that can distinguish between
them.  Intuitively, two events are exclusive when they cannot happen
simultaneously in the same run of the experiment.  For example, in the
Bell scenario depicted in Fig.~\ref{fig:chshinstdag}-b, events where we
get $a = 0$ or $a = 1$, while setting $x = 0$ for both, are exclusive,
since only one of them can happen in a single run of the experiment. On
the contrary, if the setting $x$ is different (for example $x = 1$ for
$a=0$ and $x = 0$ for $a=1$), the events will not be exclusive since a
single experimental test cannot distinguish between them. In the next
section we will provide a precise definition of exclusivity which will
allow us to extend this concept to a wide range of causal models.

Any linear constraint (like the instrumental inequalities) can be
expressed by a linear function
\begin{equation}
    I_w(p) = \sum_{\substack{a_1,\ldots,a_n\\x_1,\ldots,x_n}}
w_{\mathbf{a}, \mathbf{x}} p(a_1,\ldots,a_n|x_1,\ldots,x_n),
\end{equation}
on the probabilities of possible events. This linear function can be
embedded in an exclusivity graph by weighting the vertices in $G$ with
the $\left\{w_{\mathbf{a}, \mathbf{x}}\right\}$, so that it can be written
as a function of $G = (V, E)$ and its weights as
\begin{equation}
    I(G,w) = \sum_{v \in V}w_v p(v).
    \label{eq:linconstG}
\end{equation}
Nicely, as it will be discussed below, bounds for the maximum values
of $I_w(p)$, achievable both in the classical and quantum cases, can
be related to two well-known graph invariants \cite{cabello2014}: the
independence number $\alpha(G, w)$ and the Lovász theta $\theta(G, w)$,
respectively. In the following, we will briefly introduce these concepts
and their interconnections, a more extensive and detailed account can
be found in \cite{cabello2014,rabelo2014,acin2015}

Consider a graph $G(V, E)$ with vertex weights $w$, and $|V| = n$. We
call a \emph{characteristic labelling} for $U \subseteq V$ a vector $x_v
\in \{0,1\}^n$ such that $x_v = 1$ if $v \in U$ and $x_v = 0$ otherwise.
An \emph{independent set} or \emph{stable set} is a set $S \subset V$
such that $(u,v) \notin E$ for all $u,v \in S$.  The independence number
$\alpha(G, w)$ is defined as the maximum number of vertices (weighted
with $w$) of an independent set of $G$.  In the case of exclusivity
graphs, any characteristic labelling of a stable set, also called a
\emph{stable labelling}, represents a possible deterministic assignment of
probabilities which respects the exclusivity constraints, i.e. such that
no exclusive events can be assigned probability one at the same time.
It is also customary to define the set $\STAB(G)$ as the convex hull of
all the characteristic labellings of stable sets, such that
\begin{multline} 
    \STAB(G) = \chull (\{x : \\ x \quad \text{is a stable labelling of}\quad G \}).
    \label{eq:stab}
\end{multline}
Since stable labellings represent all the possible deterministic strategies respecting the exclusivity relations, then $\STAB(G)$ effectively includes  all the possible 
probability assignments compatible with those constraints.
Now we can define the independence number as
\begin{equation}
    \alpha(G,w) = \max\{w\cdot x: x \in \STAB(G)\}.
    \label{eq:alphastab}
\end{equation}
Thus, $\alpha(G,w)$ must correspond to the classical bound of the
inequality, since it is exactly the maximum over the convex set defined by
all the deterministic strategies respecting the exclusivity constraints.
Classical models are those described precisely by such convex set,
thus implying that
\begin{equation}
    I(G,w) \le \alpha(G,w).
    \label{eq:linconstG_alphabound}
\end{equation}
Moreover, the bound is tight since it is saturated by any deterministic
assignment corresponding to a maximal stable set.

In associating the set $STAB(G)$ with the space of the possible
probability distributions for our graph $G$, we have made the implicit
assumption that there exists a joint probability distribution for all
of our events, i.e., that even when certain settings are not chosen
by the experimenter, we can still assign (counterfactually) a value to
their probabilities.  This is the \emph{realism} assumption mentioned
above that, as implied by Bell's theorem, cannot hold true togheter with
locality for quantum systems.  In the following, we briefly introduce the
probabilistic framework used in quantum mechanics, the interested reader
can refer to classic texts such as \cite{nielsen_chuang}.  In quantum
mechanics the \emph{state} of the system, which plays a similar role
as the probability distribution for classical systems, is represented
by a vector $\Psi$ in a complex Hilbert space $\mathcal{H}$, normalized such
that $\abs{\Psi}^2 = 1$. Likewise, measurements are associated to a
set to an orthonormal basis $\{\Phi_1,\ldots,\Phi_d\}$ in the same
space $\mathcal{H}$, each associated to a possible measurement outcome
\footnote{This actually describes a particular class of measurements
called \emph{projective measurements.}}. It is also customary to
represent measurements using projection operators $E_i = \Phi_i
(\Phi_i)^T$, so that $E_i E_j = 0 \, \forall i,j$ and $\sum_i E_i =
I$. The probability associated to each outcome is defined by the
\emph{Born's rule}:
\begin{equation}
    p_i = \Psi^T E_i \Psi = \abs{\Psi \cdot \Phi_i}^2.
\end{equation}
It is known that such framework allows for a set of probability
distributions which is in general larger than the classical one.
Exclusivity relations between events (measurement outcomes) in the
quantum framework translate into orthogonality between projectors.
A quantum realization of an exclusivity graph $G(V,E)$ then consists in a
set projectors $E_i$ for each vertex $i \in V$, such that $E_i$ and $E_j$
are orthogonal each time $i$ and $j$ are connected by an edge. This
corresponds to what in graph theory is known as an \emph{orthonormal
labelling} of $G$.
An \emph{orthonormal labelling} of dimension $d$ is a map $a_v:V
\rightarrow \Real^d$ such that $a_v \cdot a_u = 0$ for all $(u, v) \in E$
and $|a_v|^2 = 1$.
Using that notion we can define the set $\TH(G)$ as
\begin{multline}
    \TH(G) = \{x: x_v = (a_v)_1 \, \text{where $a_v$ is an} \\ \text{orthonormal labelling of $G$}\}.
    \label{eq:thbody}
\end{multline}
It can be proved that this set includes all correlations permitted by quantum theory 
but in general is larger as it contains correlations beyond those achievable by quantum
mechanics \cite{almostquantum2015}.
Maximizing over $\TH(G)$ led to the Lovász theta given by
\begin{equation}
    \theta(G,w) = \max \{w\cdot x : x \in \TH(G)\},
    \label{eq:lovasztheta}
\end{equation}
which upper-bounds the maximum quantum value of $I(G,w)$ in equation
\eqref{eq:linconstG}.  Despite not being a tight bound for quantum
system in general, $\theta(G,w)$ is known to be efficiently computable,
by a semi-definite program.

This also provide a useful condition to check if a given graph $G$ (or
any of its induced subgraphs) admits a quantum violation.  Indeed using
a known result of graph theory we know $\TH(G) = \STAB(G)$ if and only
if $G$ does not contain a cycle $C_n$ with $n \ge 5$ and odd, or its
complement as induced subgraphs.  This follows directly from the so
called \emph{sandwich theorem} \cite{knuth,lovasz} and the \emph{strong
perfect graph theorem} \cite{spgth}.  The first one states that the
number $\theta(G)$ is always greater or equal the independence number
of the graph $\alpha(G)$.
\begin{equation}
    \alpha(G) \le \theta(G)
    \label{eq:sandwich_th}
\end{equation}
When equality holds in equation \eqref{eq:sandwich_th} for a graph
$G$ and all its induced subgraphs, $G$ is called \emph{perfect}.
For perfect graphs, we can exclude the existence of a quantum violation,
since $\alpha(G) = \theta(G)$.  The second theorem then gives a useful
condition to check if a graph is perfect or not. In particular it affirms
that a graph $G$ is perfect if and only if it does not contain a $n$-cycle
graph with $n\ge5$ and odd or its complement as an induced subgraph.
Besides signaling the presence of a possible quantum violation of
a classical constraint, induced cycle subgraphs are also interesting
because they give the simplest inequalities (in terms of number of
probabilities to estimate), to test this violation.

\section{EXCLUSIVITY GRAPH METHOD APPLIED TO CAUSAL MODELS}
\begin{figure}[h]
    \centering
    \includegraphics[width=.9\columnwidth]{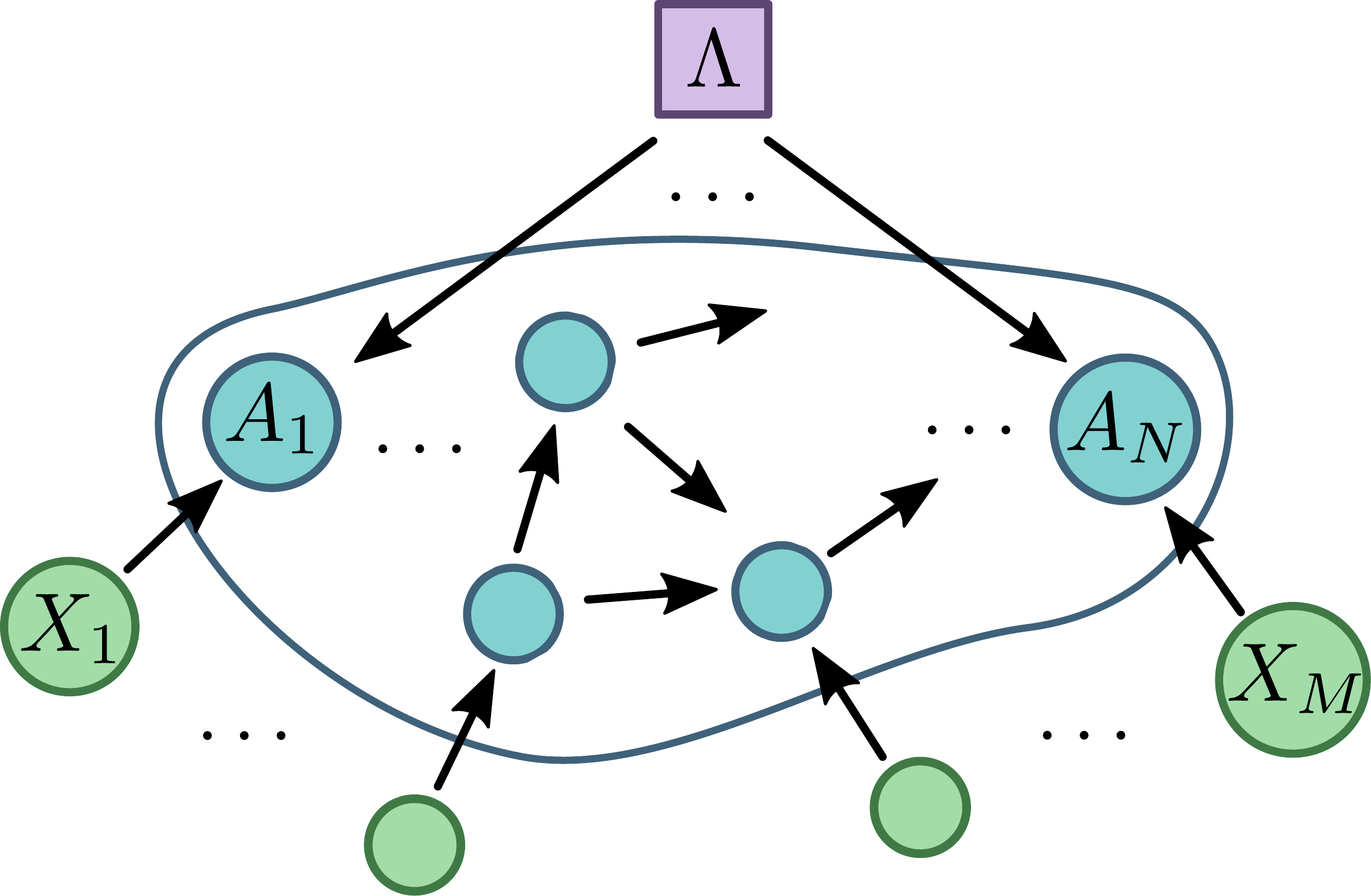}
    \caption{Represetantion of the class of causal structures to which our
        method can be applied, which are those with $k$ observable variables, $l$
        instruments and a single latent variable.}
    \label{fig:onelambda}
\end{figure}

In this section we show how the techniques presented in the
previous section can be employed to analyze a broad class of causal
models. Consider the DAG depicted in Fig~\ref{fig:onelambda}, with $N$
observable variables $A_1, \ldots, A_N$ with arbitrary causal arrows
among them, $M$ instruments $X_1, \ldots, X_M$, and a single unobservable
latent variable $\Lambda$ acting as a potential common factor for all
$A_i$s (but not for the $X_j$s).

An \emph{exclusivity} graph can be associated with such a DAG
as follows:
\begin{itemize}
    \item Nodes are associated to events like $a | x$, where $a = (a_1,\ldots,a_N)$ and  $x = (x_1,\ldots,x_M)$.
    \item Two nodes $a | x $, and $a' | x'$ are linked by an edge if
    there is at least a variable $A_i$ for which \emph{does not} exists
    any function $f_i$ such that: 
    \begin{equation}
        f_i(\pa_i) = a_i \et f_i(\pa_i') = a'_i.
    \end{equation}
    where $\pa_i$ and $\pa_i'$ represent the values taken by the parents
    of $A_i$ in the two events.
\end{itemize}

For example, referring to the Bell scenario of
Fig.~\ref{fig:chshinstdag}-b any two events $a,b,x,y$ and $a',b',x',y'$
where $x = x'$ and $a\neq a'$ (or $y = y'$ and $b \neq b'$) are exclusive,
since we would need $f(x) \neq f(x')$ even if $x = x'$ (or similarly
$g(y) \neq g(y')$ when $y = y'$).  As we will show next, considering the
particular case of the instrumental scenario \cite{pearl1995, bonet2001},
one can apply the graph-theoretical methods delineated before to its
corresponding exclusivity graph, $G$, and its induced subgraphs. This
allows to obtain instrumental inequalities and their respective quantum
and classical bounds.  To do that, we proceed as follows: First we try to
determine if the graph is perfect using the \emph{strong perfect graph
theorem}, i.e. looking for odd cycles and anticycles with more than $5$
vertices among the induced subgraphs of $G$.  If the graph is perfect,
then we know immediately that no quantum violation is possible.  If the
graph is not perfect, then we must have found some odd cycle or anticycle
$C_n$. These kinds of induced subgraphs represent our minimal candidates
for a quantum violation, cause we already know that for them $\alpha(C_n)
< \theta(C_n)$, in particular:
\begin{equation}
\begin{aligned}
    \alpha(C_n) &= \lfloor n/2 \rfloor \\ 
    \theta(C_n) &= \frac{n\cos(\pi/n)}{1+\cos(\pi/n)}
\end{aligned}
\label{eq:cycle_alpha_theta}
\end{equation}
Other candidates can be found among the induced subgraphs of $G$,
possibily with non-unitary weights $w$, which contain at least one of
those cycles/anticycles.  So any weighted subgraph $S$ that satisfies
$\alpha(G,w) < \theta(G,w)$ in the end must contain at least a unitary
weighted subgraph with this same property.  For this reason in the
following analysis we focus on cycles or on unitary weighted graphs only.

\subsection{THE INSTRUMENTAL EXCLUSIVITY GRAPH}
\begin{figure}[t]
    \centering
    \includegraphics[width=\columnwidth]{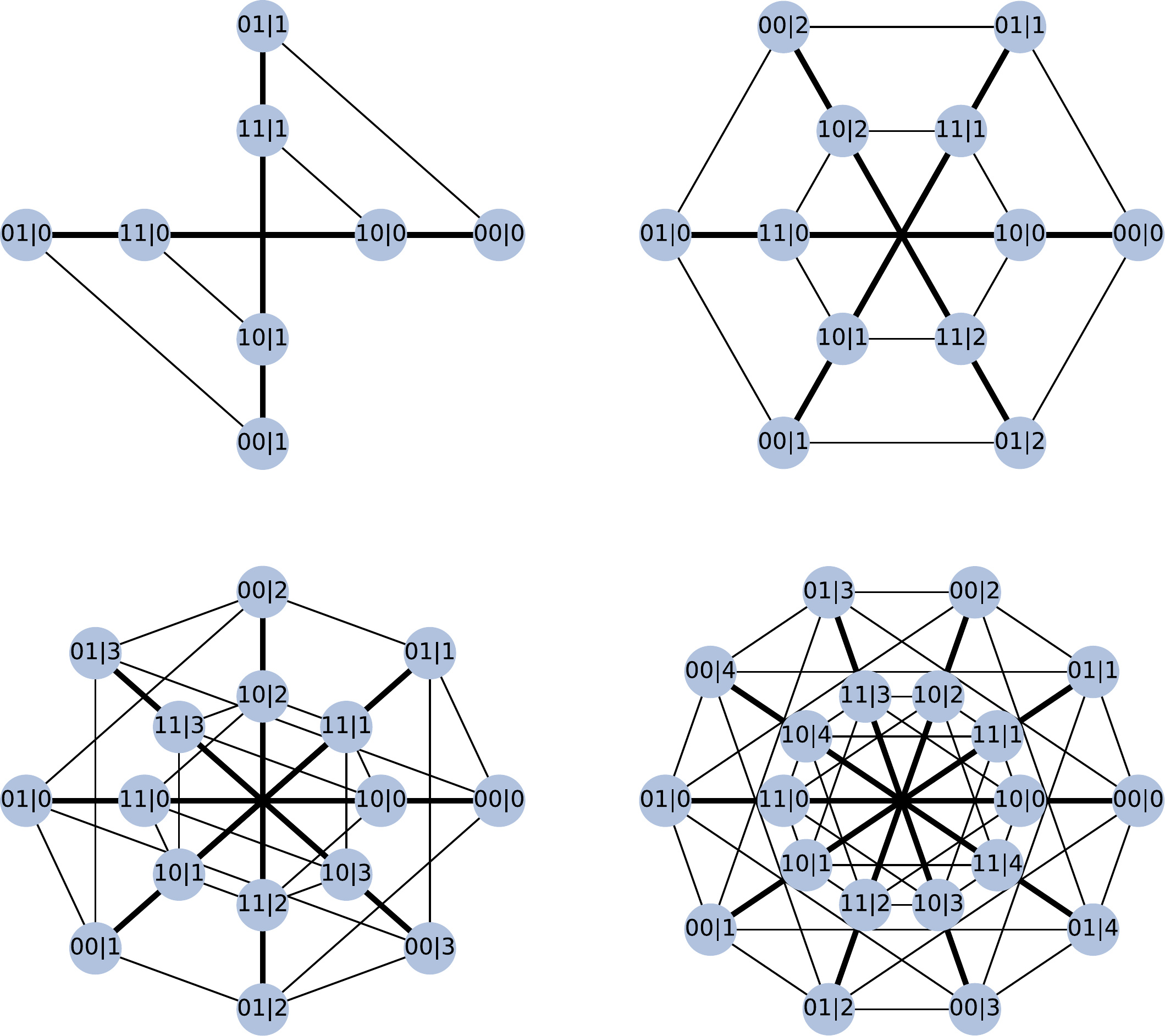}
    \caption{
    The exclusivity graph for the instrumental scenario with $m =
    n =2$ and $l=2,3,4,5$ respectively from top left to bottom right.
    To simplify the representation cliques are represented with the bold
    lines in the figure.}
    \label{fig:instrumental_exgraphs}
\end{figure}

As a first application we will restrict our attention to the instrumental
scenario in the case of dichotomic measurements ($n = m = 2$).  We denote
the probability of having outcomes $a$ and $b$ with the instrument
assuming the value $x$ as $p(ab|x)$ with $a, b \in \mathcal{A} =
\mathcal{B} = \{0,1\}$ and $x \in \mathcal{X} = \{0,\ldots,l\}$.
As explained above, the exclusivity graph for the instrumental scenario
is obtained by connecting two events $a,b,x$ and $a',b',x'$ with an edge
if we cannot find a function $f:\mathcal{X} \rightarrow \mathcal{A}$
such that:
\begin{equation}
    a = f(x) \quad\text{and}\quad a'=f(x')
\end{equation}
or a function $g:\mathcal{A} \rightarrow \mathcal{B}$ such that:
\begin{equation}
     b = g(a) \quad\text{and}\quad b'=g(a')
    \label{eq:non_exclusivity_condition}
\end{equation}
Using these rules we construct the exclusivity graphs $G_l
= (V_l, E_l)$ for various $l$, some of which are shown in
Fig.~\ref{fig:instrumental_exgraphs}, and use the methods described in
the previous sections to obtain the classical and quantum bounds for
several inequalities in the instrumental scenario.

First, consider the case $l=2$, depicted in
Fig.~\ref{fig:instrumental_exgraphs} top-left, for which Pearl's
inequality \eqref{eq:pearl_ineq} defines the only instrumental inequality.
It has been shown that this inequality does not have a quantum violation
\cite{henson2014}.  For that, general probabilistic Bayesian networks,
including classical and quantum causal models as particular cases, had
to be introduced.  In contrast, in our method it is straightforward not
only to derive the classical bound to Pearl's inequality but also show
that there is no quantum violation of the inequality.  In the case of
$l=2$ inequalities \eqref{eq:pearl_ineq} becomes:
\begin{equation}
    P(a0|x) + P(a1|x') \le 1 \, \forall a,x,x' \in \{0,1\} 
    \label{eq:pearl_ineq_222}
\end{equation}
which are just the classical constraint given by the exclusivity
conditions represented by the edges of the graph (see
Fig.~\ref{fig:instrumental_exgraphs}).  Indeed considering the
trivial induced subgraphs $S_e$ formed by only two vertices $e =
(v_1, v_2) \in E_2$, we simply have $\alpha(S_e) = 1$ from which
using equation \eqref{eq:linconstG_alphabound}, we obtain contraints in
\eqref{eq:pearl_ineq_222}.  The fact that no quantum violation is allowed
follows immediately from the fact that the corresponding  exclusivity
graph (and its complement) does not contain any odd cycle or anticycle
with more than $5$ vertices, which makes it a perfect graph, i.e. $TH(G)
= STAB(G)$. This can be easily proved in the case $l=2$ for any number
of outputs $n,m$ as shown in the supplemental materials. In this way
we can exclude the presence of any quantum violation for any scenario
where the instrument can only take two possible values ($l=2$) and an
arbitrary number of outputs for $A, B$.

\begin{figure}[t]
    \centering
    \includegraphics[width=.6\columnwidth]{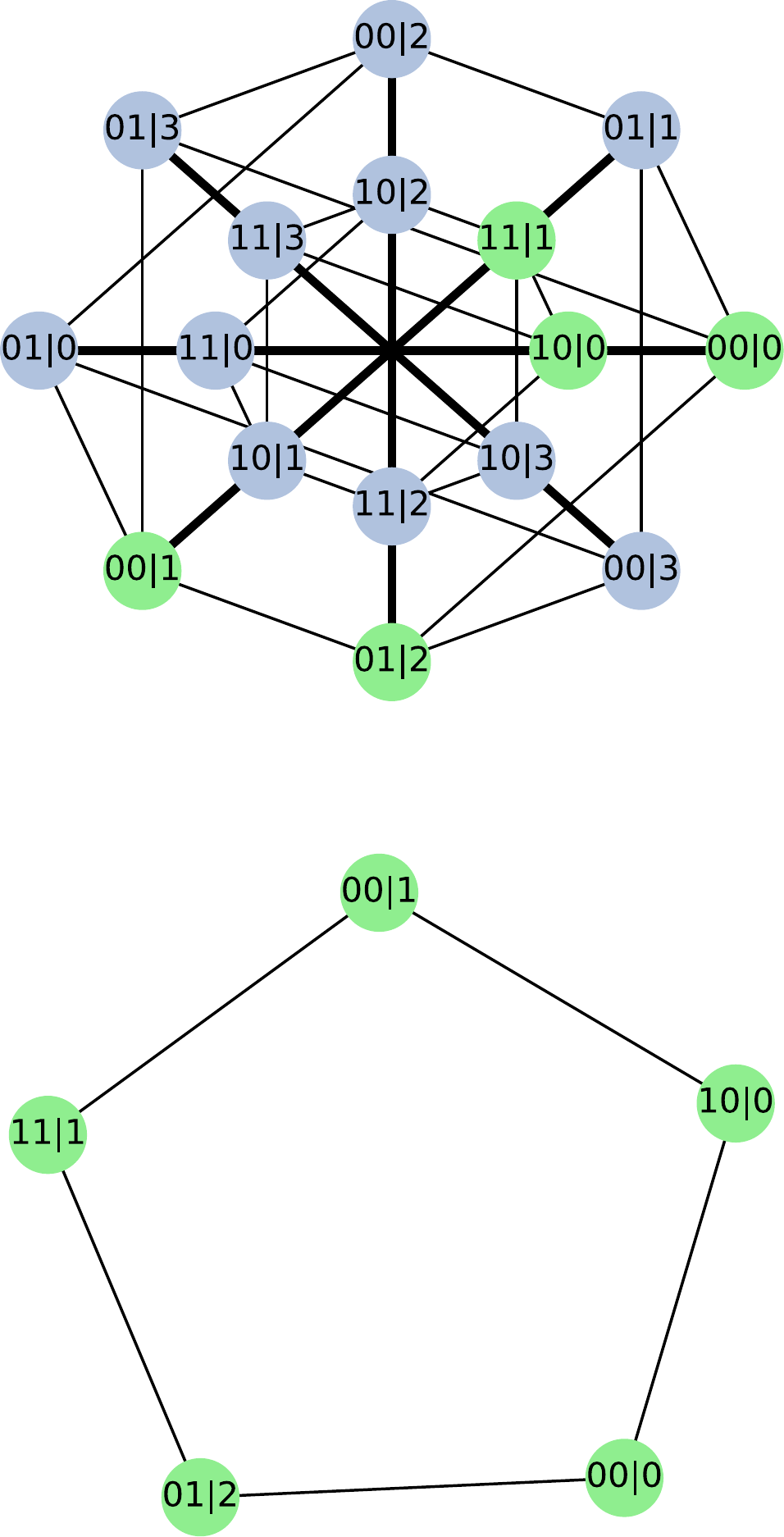}
    \caption{The exclusivity graph of the bonet inequality as an induced
    subgraph of complete one of the $(l,m,n)=(4,2,2)$ instrumental
    scenario. To simplify the representation, cliques are represented
    with the bold lines in the figure.}
    \label{fig:bonetexc}
\end{figure}

Going to higher number of outcomes for the instrument $X$ we see
that there might be a quantum violation, since the associated graph
$G_l$ has as a $C_5$ cyclic graph as induced subgraph for $l \ge 3$.
The graph $C_5$, depicted in Fig.~\ref{fig:bonetexc}, represents an
instance of Bonet's inequality \eqref{eq:bonet_ineq}, and indeed from
equation \eqref{eq:cycle_alpha_theta} we get the expected classical bound
$\alpha(C_5)=2$.  The quantum limit, as already mentioned, in general
does not saturate the bound given by $\theta(G)$, which in this case is
(from \eqref{eq:cycle_alpha_theta}) $\theta(C_n) = \sqrt(5)$.  To find a
tighter bound we can apply the technique described in \cite{rabelo2014},
which is in turn based on the so called NPA method (from M. Navascués,
S. Pironio, A. Acín) described in \cite{npa2008}.  This is a method
commonly employed in quantum information to perform optimizations
constrained in the set of quantum correlations. Since, in general,
there are no known methods to impose this exact optimization condition,
the technique works by relaxing the problem to a, virtually infinite,
hierarchy of semi-definite programs of increasing dimension, which
approximate the restriction to quantum correlations.  Applying it to
Bonet inequality we are able obtain the known result for the maximum
quantum bound, i.e $(3+\sqrt{2})/2 \approx 2.2071$ (more details can be
found in the supplemental materials).

As shown in the supplemental material, no other odd cycle besides $C_5$
is present for any $l$, that is, if we increase the cardinality of the
instrumental variable. The first $7$-cycle appears as soon as we get to
$n=m=3$ and $l=4$.
An instance of this Bonet-like inequality for the instrumental scenario can be written as:
\begin{multline}
    P(00|2) + P(02|3) + P(00|0) + P(12|0) + \\
    + P(10|1) + P(21|1) + P(22|2) \le 3 
    \label{eq:c7_instrumental433}
\end{multline}
Applying the method cited above for this inequality gives a quantum
upper bound of $q = 3.2990$ at the second order of the hierarchy, which
indicates the possibility of a quantum violation.  Similarly, numerical
analysis shows that odd cycle subgraphs with higher number of vertices
appear only if we increase the number of possible settings $l$ while also
increasing $m$, so for example $9$-cycles start to appear for $l=6, m=3$
and $11$-cycles for $l=7, m=4$.

\begin{figure}[t]
    \centering
        \includegraphics[width=\columnwidth]{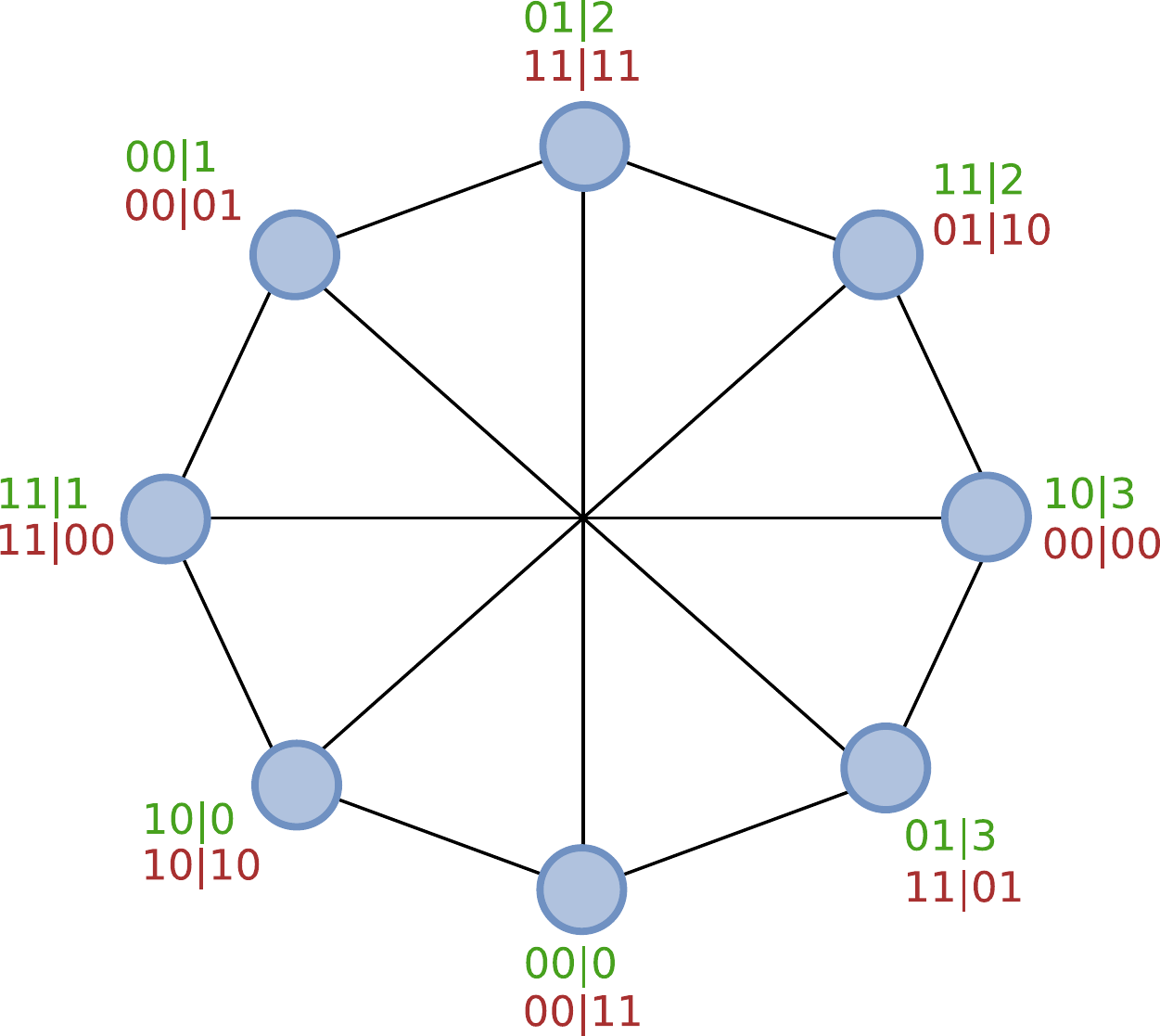}
        \caption{Exclusivity graphs for the the inequality \eqref{eq:chsh_ineq} in the Bell
        scenario (in red) and inequality \eqref{eq:422_ineq} in the instrumental
    scenario (in green).}
        \label{fig:422_exgraph}
\end{figure}

While cycles are the simplest inequalities showing quantum violation, our method
can also be employed for the analysis of different inequalities, that can be 
devised by other choices of vertices. 
For example in the instrumental scenario $(l,m,n) = (4,2,2)$ we can find by inspection the inequality: 
\begin{multline}
    P(01\vert2) + P(11\vert2) + P(10\vert3) + P(01\vert3) + \\
     + P(00\vert0) + P(10\vert0) + P(11\vert1) + P(00\vert1) \le 3
    \label{eq:422_ineq}
\end{multline}
This inequality is interesting, since it is represented by the same
exclusivity graph of the notorious CHSH inequality \cite{CHSH} for the
Bell scenario (see Fig.~\ref{fig:422_exgraph}):

\begin{multline}
 P(00\vert 00) + P(11 \vert 00) + P(00\vert 01) + P(11 \vert 01) + \\
 +P(01 \vert 10) + P(10 \vert 10) + P(00 \vert 11) + P(11 \vert 11) \leq 3
\label{eq:chsh_ineq}
\end{multline}

A well known generalization of the CHSH inequality are the so called CGLMP
(Collins, Gisin, Linden, Massar and Popescu) inequalities, introduced
in \cite{cglmp}, which are defined for any Bell scenario with 2 settings
for $X$ and $Y$ and $d$ outcomes for $A$ and $B$, and can be written as:
\begin{gather}
    I_d^{\mathrm{CGLMP}} = \sum_{k=0}^{d-1} (d-1-k) S^d_k \le 3 (d-1)\\ \nonumber
    \text{where}\\
    \quad S^d_k = \sum_b (P(b+k,b|00) + P(b+k, b | 11)) + \\ \nonumber
     + \sum_a (P(a, a+k+1|10) + P(a, a+k | 01))
    \label{eq:GCLMP}
\end{gather}
where the sums $a+k, a+k+1$ and $b+k$ are modulo $d$.
Using the exclusivity graph method we can find that each of the $S^d_k$ is
classically constrained by:
\begin{itemize}
    \item $\alpha(G^d_k) = 4$ if $k$ and $d$ satisfy $4k+1=n d$, for some
        integer $n$.
    \item $\alpha(G^d_k) = 3$ in the other cases.
\end{itemize}
Indeed the graphs $G^d_k$ relative to the $k$ all share the same
structure: there are four cliques, one for each setting $x,y \in \{0,1\}$,
and any vertex in each clique is connected to every other vertex in the
adjacent clique, except for one.  For example $P(b+k,b|00)$ is connected
to any node belonging to the $(0,1)$ and the $(1,0)$ cliques, except for
$P(a,a+k|01)$ and $P(a',a'+k+1|10)$ where $b+k = a$ and $a'+k+1 = b$.
Clearly a maximal independent set cannot contain more than $4$ vertices
(one for each clique).  Moreover to be an independent set, a set of four
nodes $\{P(b+k,b|00), P(b'+k,b'|11), P(a,a+k|01), P(a',a;+k+1|01)\}$
must satisfy the conditions:
\begin{equation}
    \left\{
        \begin{aligned}
            &b + k = a \\
            &a + k = b' \\
            &b' + k = a' \\
            &a' + k + 1 = b 
        \end{aligned}
    \right.
    \label{eq:four_indipset_condition}
\end{equation}
where the sums are all modulo $d$.
From this follows directly that $4k+1 = 0$. 

To obtain the quantum bounds we can apply the same method discussed above.
The results for some $S_k$ inequalities are shown in Table~\ref{tab:gclmps}.
\begin{table}
    \centering
    \begin{tabular}{ccccc}
        $d$ & $k$ & $\alpha(G^d_k)$ & $\theta(G^d_k)$ & NPA \\
        \toprule
         3 & 0 & 3 & 3.464 & 3.333 \\
         3 & 1 & 3 & 3.464 & 3.333 \\
         4 & 0 & 3 & 3.414 & 3.307 \\
         4 & 1 & 3 & 3.414 & 3.307 \\
         5 & 0 & 3 & 3.431 & 3.294 \\
         5 & 1 & 4 & 3.999 & 3.999 \\
    \end{tabular}
    \caption{Considering the inequality $S^d_k$ for different values of $k$ and $d$, the table above shows the independence number $\alpha(G^d_k)$, the Lov\'asz theta $\theta(G^d_k)$ and the NPA bound computed up to the second order of the hierarchy.}
    \label{tab:gclmps}
\end{table}

Interestingly, except for the case $(l,m,n) = (4,2,2)$, inequalities with the
same structure do not seem to arise in the instrumental case, which suggests
that the apparent similarity noticed in \cite{himbeeck2018} between the two scenarios, Bell and the
instrumental, only appears for specific number of inputs and outputs.

\section{DISCUSSION}
In this paper, we have proposed an unifying formalism to analyze classical and
quantum correlations arising in a broad class of causal structures. It is based
on a graph-theoretical formalism originally introduced in the field of quantum
information \cite{cabello2014,rabelo2014,acin2015}. In particular, we consider
the application of this formalism to analyze instrumental tests
\cite{pearl1995}. As we show, the probabilities arising in such experiments can
be encoded in a exclusivity graph and from there it follows that the classical
and quantum bounds respected by instrumental inequalities are related to two
graphs invariants: the independence number, $\alpha$, and Lov\'asz $\theta$,
respectively.

Apart from the fundamental relevance of bridging the fields of quantum
information and causal inference, our approach is also shown to be of practical
use. We not only re-derived, in an easy manner, previous results in the literature,
we also manage to generalize them. For instance, we prove the inequalities
associated with an instrument assuming only two possible values do not have a
quantum violation (irrespectively of the number of outcomes), thus generalizing
the results in \cite{henson2014}. As well, we prove that if the number of
outcomes is fixed to two (the instrument now assuming any cardinality), there are no
other inequalities other than the original Bonet's inequality \cite{bonet2001} arising from a n-cycle graph. Following that, we have shown how new
instrumental inequalities associated with n-cycles of increasing n can be
obtained by increasing the possible values of both the instrument and the
outcomes.

The graph approach also constitutes a valuable tool to study similarities among
different scenarios and inspect whether, in the quantum realm, they could be
able to detect stronger forms of non-locality. For example, from the graph
perspective, the instrumental scenario and the well-known Bell scenario shows
similarities only for specific number of inputs/outputs. For example, the CHSH
scenario \cite{CHSH} and the $(l,m,n)=(4,2,2)$ instrumental scenario are graph
equivalent, however, this equivalence does not hold any longer when the outcome
variables assume an increasing number of possible values. Given the fundamental
importance of the instrumental scenario in causal inference and the increasing
attention it has been receiving in quantum information (particularly in
applications as randomness generation) we hope these results will strength the
connections between both fields and motivate further applications of the
graph-theoretical approach within causality.

\subsubsection*{Acknowledgements}
We acknowledge support from John Templeton Foundation via the grant Q-CAUSAL
n$^{\circ}$61084 (the opinions expressed in this publication are those of the
authors and do not necessarily  the views of the John Templeton
Foundation). RC acknowledges the Brazilian ministries MEC and MCTIC, funding
agency CNPq (PQ grants No. 307172/2017-1 and No 406574/2018-9 and INCT-IQ) and
the Serrapilheira Institute (grant number Serra-1708-15763).

\clearpage
\section*{SUPPLEMENTAL MATERIAL}

\subsection*{Building the exclusivity graph from DAG}
In the following we describe in more details how to get from the DAG
(Directed Acyclic Graph) representation of a causal model to the one
for exclusivity graph.
Starting from a generic causal model described by a DAG $D$, with
$N$ random variables $O_D = \{A_1,\ldots,A_N\}$ and $M$ instruments
$I_D = \{X_1,\ldots,X_M\}$, the exclusivity graph $G = (V, E)$ can be
constructed, for example, using a simple breadth-first graph exploring
algorithm.
The procedure, described in algorithm \ref{alg:bfgb}, requires the DAG $D$
and the list $V$ of vertices to be explored, since we can be interested
in building the graph only for a subset of events.

\begin{algorithm*}
\caption{Breadth-first graph exploration}
\label{alg:bfgb}
\begin{algorithmic}[1]
\Function{build graph}{$V, D$} 
\State $E \gets \emptyset$
\While{$V \neq \emptyset$}
    \State $\Call{insert}{Q, V_1}$ \Comment{Initialize the queue with the first element of $V$}
    \State $\Call{delete}{V, V_1}$
    \While{$Q \neq \emptyset$}
        \State $v \gets Q_1$ 
        \State $\Call{delete}{Q, Q_1}$
        \For{$u \in V$}
            \If{\Call{exclusive}{$u$, $v$}}
                \State \Call{insert}{$E, (v, u)$}
                \State \Call{insert}{$Q, u$}
                \State \Call{delete}{$V, u$} \Comment{Visited nodes are removed from $V$}
        \EndIf
        \EndFor
    \EndWhile
\EndWhile
\State \Return E
\EndFunction
\end{algorithmic}
\vspace{1em}
As in the main text, here $a, a'$ stand for the value of the outcome of the
variable $A$ in the events $v, v'$, while $p_a, p_{a'}$ stand for the values
of the parent nodes of $A$ in $D$, $\PA (A)$.

\begin{algorithmic}[1]
\Function{exclusive}{$v, v', D$} 
    \State $n \gets \text{true}$
    \For{$A \in O_D$} 
        \State $n \gets n \land \left(p_a \neq p_{a'} \lor (p_a = p_{a'} \land a = a') \right)$
    \EndFor
    \State \Return $\neg n$
\EndFunction
\end{algorithmic}
\end{algorithm*}

\subsection*{Edge colored multigraph technique for approximating the quantum
bound}
\begin{figure}[t]
    \centering
    \includegraphics[width=.7\columnwidth]{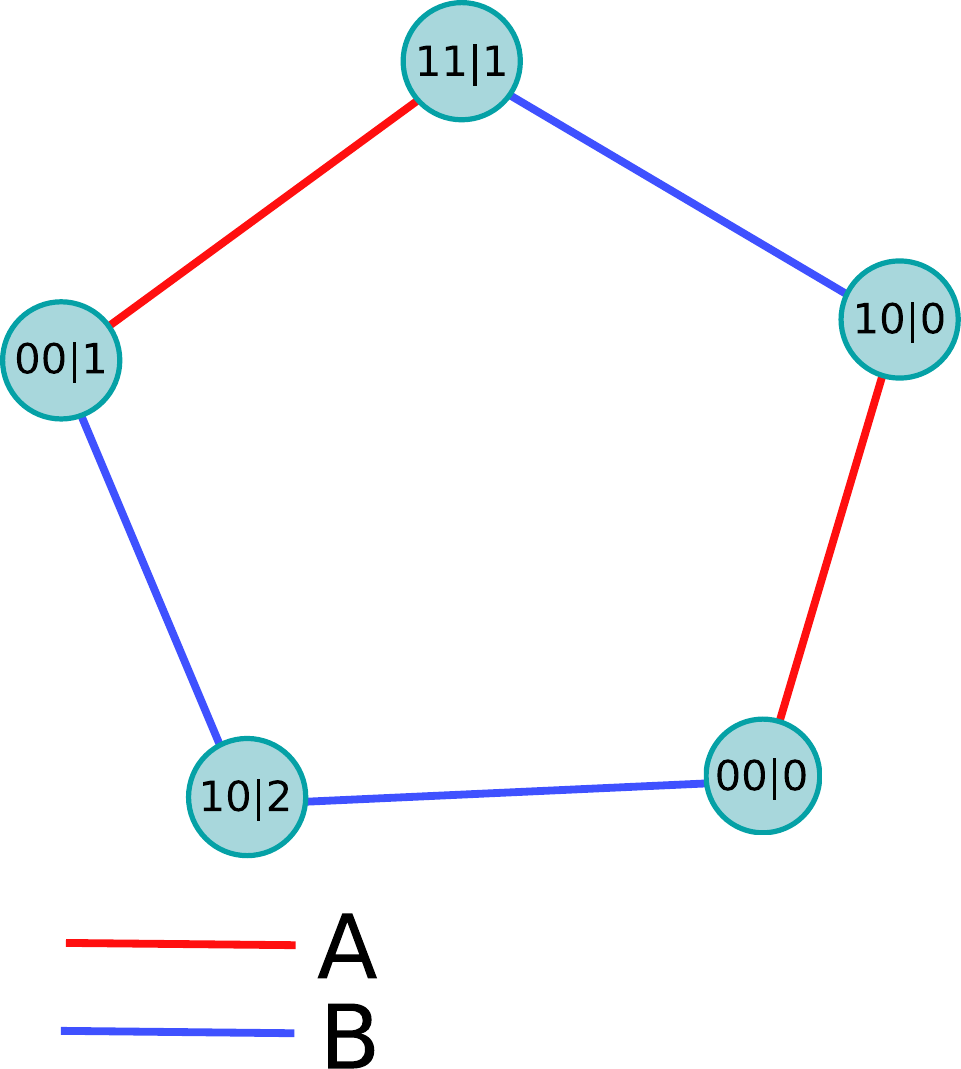}
    \caption{Edge colored exclusivity graph representation of the Bonet
        inequality. Exclusivity constraints for the party $A$ and $B$ are
    represented by red lines and blue lines respectively.}
    \label{fig:instrumental_multigraph}
\end{figure}

The Lov\'asz theta of a graph, despite being efficiently computable,
only gives an upper bound to the maximal quantum bound, since it ignores
the additional constraints arising from the presence of different random
variables $A_i$.
Indeed the quantum bound is influenced not only by the exclusivity relations between the
possible events in our scenario, but also on how those relations are derived from the
variables $A_1,\ldots,A_N$.

To obtain a better approximation for the quantum bound we can follow
the technique presented in \cite{rabelo2014}.
This method consists in introducing an edge coloring in the exclusivity graph. 
This edge coloring encodes the information of which of the $A_i$s is involved in the exclusivity constraints under consideration.
In practice this corresponds to constructing an exclusivity graph $G_i$ for each
$A_i$. The resulting object is called a \emph{multigraph}. 
Having defined a multigraph $G = {G_1, \ldots, G_N}$ for a given scenario the
quantum bound is defined by the quantity:
\begin{equation}
    \vartheta(G) = \max_{v} \sum_{i \in V} |v \cdot a^1_i \otimes \dots \otimes a^n_i|^2
    \label{eq:multigraph_lovazs}
\end{equation}
where $\{a^j_i\}$ is an orthonormal labelling for $G_j$ and $V$ is the set
of vertices of $G$. This quantity, which can be seen as a generalization
of the Lov\'asz theta, is in general not efficiently computable, but,
as described in \cite{rabelo2014}, can be arbitrarily approximated by
a hierarchy of semi-definite programs\cite{npa2008}.

For example, in the case of the pentagon in the instrumental scenario we have two colors, and thus two graph $G_A$ and $G_B$, corresponding to variables $A$ and $B$ respectively, as shown in Fig.~\ref{fig:instrumental_multigraph}.
Applying the technique described above to this scenario yields a quantum
bound of $2.2071$, reproducing the known value for the quantum bound of the Bonet inequality given by $(3+\sqrt{2})/2$.

\section*{There are no quantum violation for instrumental scenarios with $l=2$
settings.}
\begin{figure}[t]
    \centering
    \includegraphics[width=.7\columnwidth]{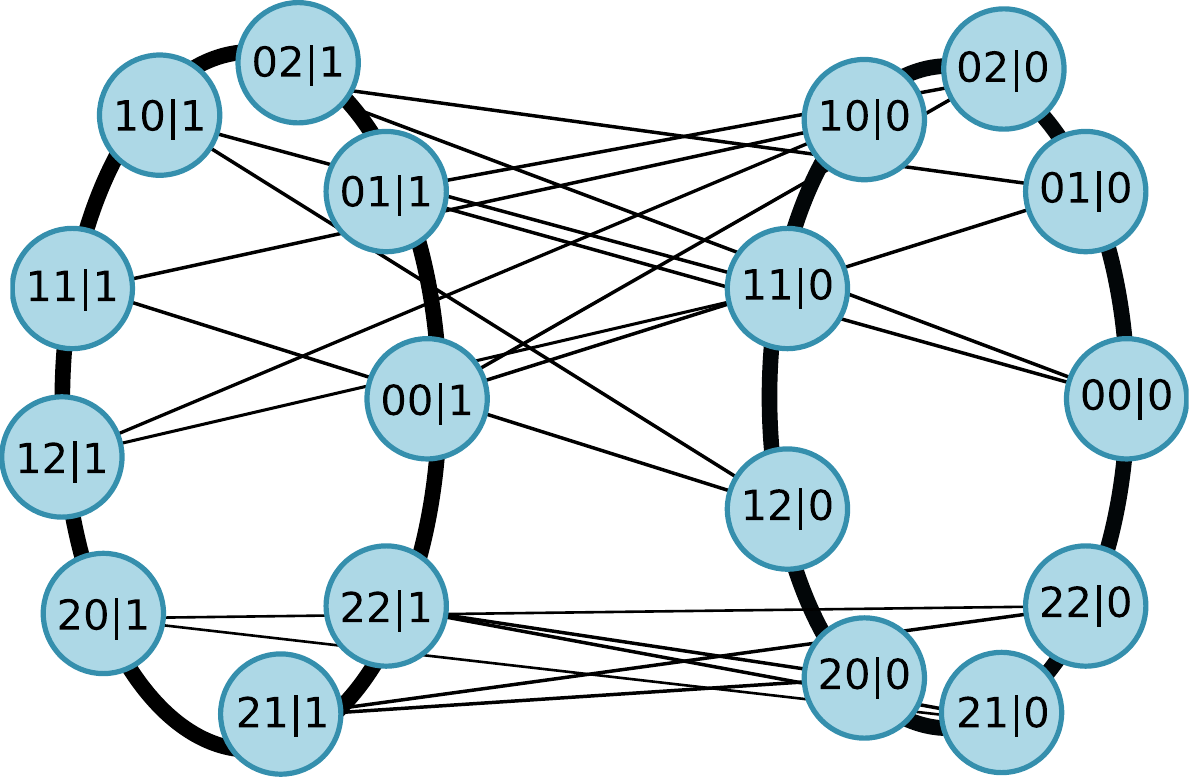}
    \caption{Exclusivity graph for the instrumental scenario $233$, showing the
        impossibility of having cycles with more than $5$ vertices. To simplify the figure cliques are
    represented by bold lines between vertices.}
    \label{fig:2mn_nocycle_proof}
\end{figure}

It is easy to see that no quantum violation is possible for instrumental scenario
with $l=2$ possible settings for the instrumental variable $X$.
This reduces to proving that there are no odd $n$-cycles nor $n$-anticycles as induced
subgraphs in the corresponding exclusivity graph, with $n\ge5$.
To see this we can notice that any such graph is composed by two cliques (see
for example Fig.~\ref{fig:2mn_nocycle_proof}), corresponding to the events with $x=0$ and
$x=1$.
Any $n$-cycle with at least $5$ vertices must then have at least $3$ mutually
connected vertices belonging to the same $x$, so they can never form a
cycle-graph.
Similarly we can show that there cannot be any induced odd anti-cycle with $5$ or more
vertices.

\subsection*{There are no cycles $C_n$ with $n \ge 7$ in the $l22$ instrumental scenario.}
In the following we prove that there cannot be a odd anti-cycle with more than
$5$ vertices in the exclusivity graph associated to an instrumental scenario of
the type $l22$.

\begin{figure}[h]
    \centering
    \includegraphics[width=.6\columnwidth]{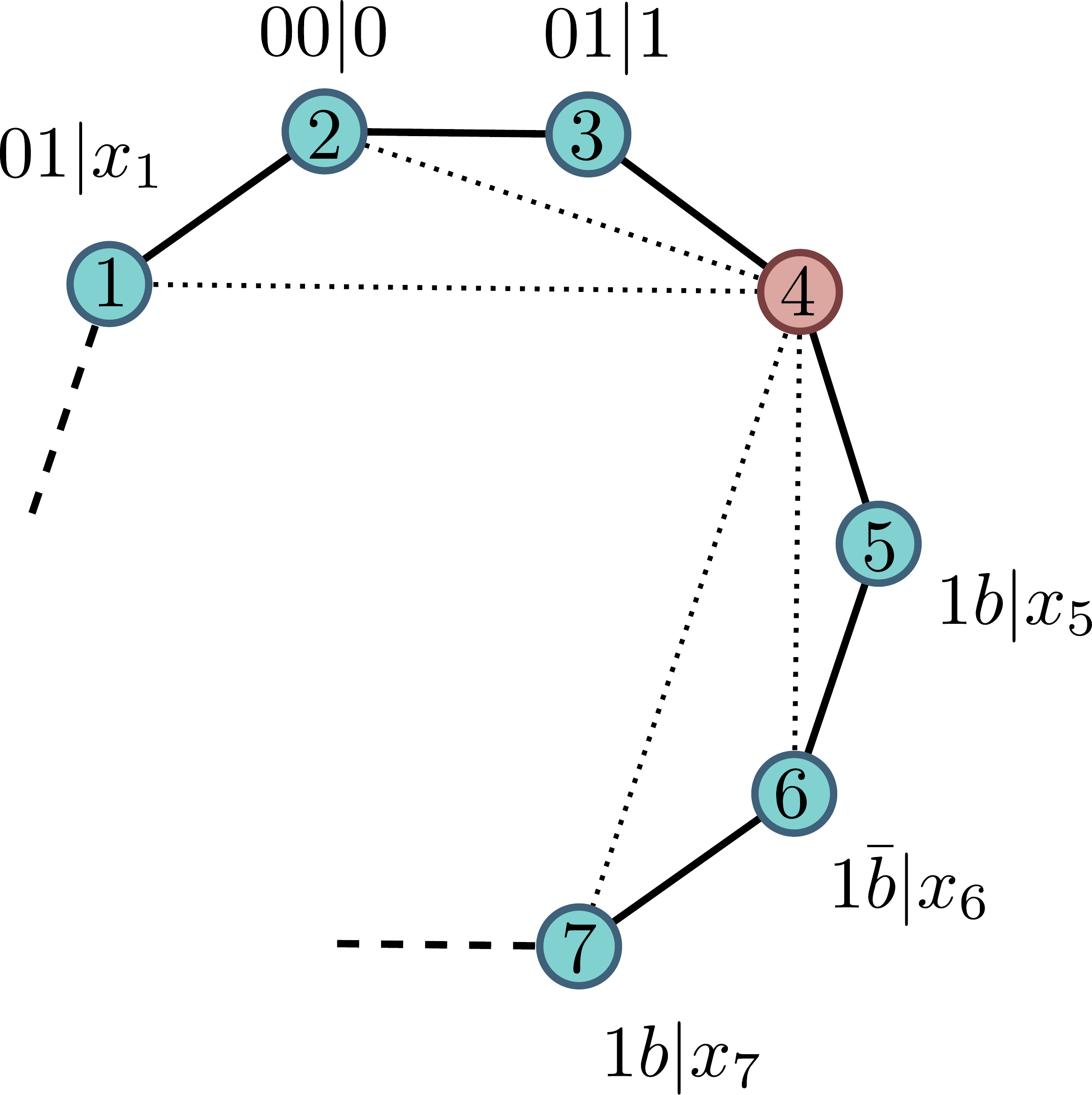}
    \caption{Proof of the impossibility of having cycles with $7$ nodes or more
    in the $d22$ scenario.}
    \label{fig:cycle_graph_proof}
\end{figure}

Two different events $ab|x$ and $a'b'|x'$, are exclusive if one of these two conditions is true:
\begin{enumerate}
    \item $x=x'$.\label{en:rule1}
    \item $a=a'$ and $b \neq b'$.\label{en:rule2}
\end{enumerate}
Suppose we have a cycle $C_n$ with $n \ge 7$, as in fig.~\ref{fig:cycle_graph_proof},
and consider that node $2$ in this graph corresponds to an event which we can arbitrarily identify as $00|0$.
Among its neighbors $1$ and $3$, one will necessarily need to satisfy rule
\ref{en:rule2} (they cannot both satisfy rule \ref{en:rule1} or the three nodes
would be a clique.
So without loss of generality we can assign the event $01|1$ to $3$.
Since nodes $5,6,7$ must not satisfy rule \ref{en:rule2} with both $2$ and $3$, then they must have $a = 1$.
Moreover $7$ and $5$ must have the same $b$, different from $6$. In the same way $1$ must not satisfy rule \ref{en:rule2} with $6,5$ and $3$, so it
needs to have $a=0$ and $b=1$. At this point, since we only have values
$\{0,1\}$ for $a$, we cannot avoid node $4$ to be linked to one of the nodes
$1,2,6,7$. Thus, the corresponding graph cannot be a cycle.

\end{document}